\documentstyle[aps,12pt]{revtex}
\begin{document}
\author{James W. Dufty}
\address{Department of Physics, University of Florida, Gainesville, FL\\
32611, USA}
\author{J. Javier Brey}
\address{F\'{\i}sica Te\'{o}rica, Universidad de Sevilla, E-41080, Sevilla,\\
Spain}
\title{Green-Kubo Expressions for a Granular Gas}
\date{\today }
\maketitle

\begin{abstract}
The transport coefficients for a gas of smooth, inelastic hard spheres are
obtained from the Boltzmann equation in the form of Green-Kubo relations.
The associated time correlation functions are not simply those constructed
from the fluxes of conserved densities. Instead, fluxes constructed from the
reference local homogeneous distribution occur as well. The analysis exposes
potential problems associated with a straightforward extension of linear
response methods to granular systems.
\end{abstract}

\pacs{KEY WORDS: Granular flow; kinetic theory; Green-Kubo; hydrodynamics.}

\section{Introduction}

The derivation of macroscopic equations (e.g., hydrodynamics) and associated
transport coefficients from kinetic theory is limited in practice to weakly
coupled systems (low density gases, ideal plasmas, anharmonic crystals). The
application of formal methods from nonequilibrium statistical mechanics to
this problem forty years ago provided the generality missing in kinetic
theory \cite{mclennan89}. These methods lead to the expected macroscopic
dynamics with formally exact expressions for the transport coefficients
known as Green-Kubo (GK) expressions. The latter are time integrals of
equilibrium time correlation functions for the fluxes associated with
conserved densities. The simplest example is the diffusion coefficient $D$
for an impurity in a simple fluid 
\begin{equation}
D=\lim_{t\rightarrow \infty ,\Omega \rightarrow \infty }\frac{1}{d}%
\int_{0}^{t}dt^{\prime }\left\langle {\bf v}(t)\cdot {\bf v}\right\rangle ,
\label{1.1}
\end{equation}%
where $d$ is the dimension of the system, ${\bf v}$ is the impurity
velocity, and the brackets denote an equilibrium ensemble average. Similar
expressions apply for the viscosities, in terms of the autocorrelation
function for the momentum flux, and the thermal conductivity, in terms of
the autocorrelation function for the energy flux. The GK expressions provide
an exact starting point for analysis and modeling of strongly coupled
systems. In particular, they have stimulated extensive studies of transport
via molecular dynamics simulation of these equilibrium time correlation
functions. Much of the present knowledge about transport in strongly coupled
systems (dense gases, liquids, plasmas, solids) derives from analysis of
appropriate GK expressions.

Currently, there is great interest in the foundations of a fluid dynamics
for granular media. Kinetic theory and statistical mechanical methods have
been applied to idealized granular fluids comprised of smooth hard spheres
interacting with inelastic collisions \cite%
{brey97statmech,dufty00granada,brey01rev,goldhirsh01,vanNoije01sm}. The
derivation of the corresponding Navier-Stokes level equations from kinetic
theory at low and moderate densities has been given recently, with transport
coefficients expressed as functions of the restitution coefficient \cite%
{brey98boltz,garzo00enskog,garzo01mix}. The results are in good agreement
with selected tests using molecular dynamics and Monte Carlo simulations %
\cite{brey01rev}. It is tempting to expect that some form of the more
general GK expressions should apply for granular media as well \cite%
{dufty00granada,goldhirsh00GK}. The primary difference from normal fluids is
the absence of a reference stable, stationary Gibbs state in terms of which
the time correlation functions would be defined. Instead, the corresponding
homogeneous state for an undriven system has a time dependence due to loss
of energy on collisions (``cooling''). This is referred to as the
homogeneous cooling state (HCS) \cite{brey96hcs,vanNoije98hcs}. Recently,
analysis of impurity diffusion in a granular fluid has led to a GK
expression similar to (\ref{1.1}) with the velocity autocorrelation function
defined in the HCS \cite{dufty02}. A related analysis of impurity mobility
also gives a GK expression for the mobility coefficient \cite{dufty01mob},
although not simply in terms of the velocity autocorrelation function as is
the case for normal fluids. The difference is due to the replacement of the
Gibbs state by the HCS. This is an indication that translation of linear
response methods for normal fluids to granular fluids requires some care.

The objective here is to derive GK expressions for a granular gas based on
the Boltzmann equation. This may seem redundant since the transport
coefficients are already known by other methods \cite{brey98boltz}. However,
the analysis exposes the form of the GK expressions that should be expected
from more general nonequilibrium statistical mechanical methods. As with the
mobility, it is found that the time correlation functions for the transport
coefficients are not simply those for the conserved densities. Instead, they
are generated from the HCS distribution in a manner required by internal
consistency for the solution to the Boltzmann equation. The analysis also
shows that the HCS correlation functions can be put in the form of
stationary state averages using appropriate dimensionless variables,
including collision number instead of real time.

It is a pleasure to dedicate this work to Bob Dorfman, colleague and mentor,
who has taught us all so much about hard sphere transport (among many other
subjects).

\section{Boltzmann equation and hydrodynamics}

The simplest model for a granular fluid is a system of $N$ smooth hard
spheres or disks at low density interacting via inelastic collisions. The
collisions are characterized by a normal restitution coefficient $\alpha
\leq 1$, where $\alpha =1$ is the elastic limit. An accurate description of
this gas can be obtained from a formal density expansion of the BBGKY
hierarchy, just as for normal gases \cite{vanNoije01sm,dufty02trieste}. To
leading order in the density, the single particle distribution function $f(%
{\bf r},{\bf v},t)$ obeys the Boltzmann equation, with collisions modified
to account for the inelasticity,

\begin{equation}
\left( \partial _{t}+{\bf v\cdot \nabla }\right) f=J[f,f].  \label{2.1}
\end{equation}
The detailed form of the inelastic collision operator $J$ is not required
for the discussion here, beyond the properties required for the macroscopic
balance equations: 
\begin{equation}
\int d{\bf v} \left( 
\begin{array}{c}
1 \\ 
{\bf v} \\ 
\frac{1}{2} mv^{2}%
\end{array}
\right) J[f,f]=\left( 
\begin{array}{c}
0 \\ 
{\bf 0} \\ 
-\frac{dp}{2} \zeta [f]%
\end{array}
\right).  \label{2.2}
\end{equation}
Here $p=nT$ is the low density pressure, $n$ being the number density and $T$
the granular temperature (with Boltzman's constant set equal to unity). The
functional $\zeta \lbrack f]$ is the ``cooling rate'', as will become
apparent below. The number density and temperature, as well as the flow
velocity ${\bf u}$ are defined in the usual way 
\begin{equation}
\left( 
\begin{array}{c}
n \\ 
n{\bf u} \\ 
\frac{d}{2}nT%
\end{array}
\right) =\int d{\bf v}\left( 
\begin{array}{c}
1 \\ 
{\bf v} \\ 
\frac{1}{2}mV^{2}%
\end{array}
\right) f,  \label{2.3}
\end{equation}
with ${\bf V}={\bf v}-{\bf u}$, the velocity relative to the local flow. The
balance equations for these fields follow directly from moments of the
Boltzmann equation using Eq.\ (\ref{2.2}), 
\begin{equation}
D_{t}n+n{\bf \nabla \cdot U}=0,  \label{2.4}
\end{equation}
\begin{equation}
D_{t}u_{i}+(mn)^{-1}\nabla _{j}P_{ij}=0,  \label{2.5}
\end{equation}
\begin{equation}
D_{t}T+\frac{2}{dn}\left( P_{ij}\nabla _{j}u_{i}+{\bf \nabla \cdot q}\right)
+T\zeta =0,  \label{2.6}
\end{equation}
where $D_{t}=\partial _{t}+{\bf u} \cdot \nabla $ is the material
derivative. The pressure tensor $P_{ij}$ and heat flux ${\bf q}$ are linear
functionals of $f$ given by 
\begin{equation}
P_{ij}[{\bf r},t|f]=p({\bf r},t)\delta _{ij}+\int d{\bf v}D_{ij}({\bf V})f(%
{\bf r},{\bf v},t),\ \quad  \label{2.7}
\end{equation}
\begin{equation}
{\bf q}[{\bf r},t|f]=\int d{\bf v}\,{\bf S}({\bf V})f({\bf r},{\bf v},t).
\label{2.8}
\end{equation}
Finally, the functions $D_{ij}({\bf V})$ and ${\bf S}({\bf V})$
characterizing the irreversible parts of the fluxes are 
\begin{equation}
D_{ij}({\bf V})\equiv m(V_{i}V_{j}-\frac{1}{d}V^{2}\delta _{ij}),\ \quad 
{\bf S}({\bf V})\equiv \left( \frac{m}{2}V^{2}-\frac{d+2}{2}T\right) {\bf V}.
\label{2.9}
\end{equation}

Clearly, the balance equations are not closed until the fluxes are expressed
as functionals of the hydrodynamic fields $n,$ ${\bf u}$, and $T$. This can
be accomplished if a solution to the Boltzmann equation can be obtained in
``normal form'', 
\begin{equation}
f({\bf r},{\bf v},t)=f\left( {\bf v}|\{y_{\beta }\}\right) ,  \label{2.10}
\end{equation}
\begin{equation}
\{ y_{\beta }({\bf r},t)\} \longleftrightarrow \left\{ n({\bf r},t), {\bf u}(%
{\bf r},t),T({\bf r},t) \right\}.  \label{2.10a}
\end{equation}
The notation $f\left( {\bf v}|\{y_{\beta }\} \right) $ means that the
distribution is a functional of the hydrodynamic fields $y_{\beta }({\bf r}%
,t)$, and that its space and time dependence occurs only through that of
these fields. If such a solution can be found, its use in Eqs.\ (\ref{2.7})
and (\ref{2.8}) gives the fluxes as functionals of the fields. Such fluxes
are then referred to as constitutive equations. The constitutive equations
together with the exact macroscopic balance equations become a closed set of
hydrodynamic equations. The origin of a hydrodynamic description in this
context therefore is traced directly to the existence of a normal solution
to the Boltzmann equation on some length and time scales. In practice, the
functional form of this solution is constructed in some well-defined
approximation, leading to corresponding approximate constitutive and
hydrodynamic equations. The approximation of interest here is small spatial
variations of the hydrodynamic fields over distances of the order of the
mean free path. For gases with elastic collisions, the Chapman-Enskog method
generates the normal solution perturbatively, and to leading order the
constitutive equations are Newton's viscosity law and Fourier's heat law %
\cite{dorfman}. The hydrodynamic equations become the Navier-Stokes
equations. Application of this method to granular gases leads to similar
results \cite{brey98boltz}. The analysis is modified here to obtain the same
results in an equivalent form with the transport coefficients represented by
GK expressions.

\section{Small gradient solution}

The first step in constructing the desired solution to the Boltzmann
equation is to express it in terms of a reference distribution $f^{(0)}$, 
\begin{equation}
f({\bf r},{\bf v},t)=f^{(0)}({\bf v}|\{y_{\beta }\})+f^{(1)}({\bf r},{\bf v}%
,t).  \label{2.11}
\end{equation}%
The reference distribution is restricted to be normal and to give the exact
moments of Eq.\ (\ref{2.3}), i.e. 
\begin{equation}
\left( 
\begin{array}{c}
n \\ 
n{\bf u} \\ 
\frac{d}{2}nT%
\end{array}%
\right) =\int d{\bf v}\left( 
\begin{array}{c}
1 \\ 
{\bf v} \\ 
\frac{1}{2}mv^{2}%
\end{array}%
\right) f^{(0)}.  \label{2.12}
\end{equation}%
The Boltzmann equation then becomes 
\begin{equation}
\left( \partial _{t}+{\bf v\cdot \nabla +}\overline{L}\right)
f^{(1)}-J[f^{(1)},f^{(1)}]=J[f^{(0)},f^{(0)}]-\left( D_{t}+{\bf V\cdot
\nabla }\right) f^{(0)},  \label{2.13}
\end{equation}%
with the definition 
\begin{equation}
\overline{L}f^{(1)}=-J[f^{(1)},f^{(0)}]-J[f^{(0)},f^{(1)}].  \label{2.13a}
\end{equation}%
The left side of Eq.\ (\ref{2.13}) generates the dynamics of $f^{(1)}$. The
solution of interest is such that $f^{(1)}$ should be proportional to the
gradients of the hydrodynamic fields, since $f^{(0)}$ provides their local
values through the constraint (\ref{2.12}). This requires that the right
side of (\ref{2.13}) be proportional to the gradients of the fields.
Evaluation of $\left( D_{t}+{\bf V\cdot \nabla }\right) f^{(0)}$ using the
macroscopic balance equations gives terms proportional to the gradients,
except for the contribution from the cooling rate in (\ref{2.6}). In
addition, the nonvanishing contribution $J[f^{(0)},f^{(0)}]$ also is not of
first order in the gradients. Therefore, the reference distribution is
finally characterized by the condition that these non-gradient terms should
vanish 
\begin{equation}
J[f^{(0)},f^{(0)}]+T\zeta ^{(0)}\frac{\partial f^{(0)}}{\partial T}=0,\quad
\zeta ^{(0)}=\zeta \lbrack f^{(0)}].  \label{2.13aa}
\end{equation}%
This is an essential point in the analysis, and it is the origin of
differences from the case of elastic collisions, as is discussed below. The
distribution function $f^{(0)}$ is the analogue of the local Maxwellian for
the case of elastic collisions, but the solution to Eq.\ (\ref{2.13aa}) for $%
\alpha <1$ is not the Maxwellian. Symmetry and dimensional analysis requires
that $f^{(0)}$ have the scaling form 
\begin{equation}
f^{(0)}({\bf v}|\{y_{\beta }\})=nv_{0}^{-d}(t)f^{(0)\ast }\left( {\bf V}%
^{\ast }\right) ,\hspace{0.3in}{\bf V}^{\ast }={\bf V}/v_{0}(t),
\label{2.15}
\end{equation}%
where $v_{0}(t)=\left[ 2T\left( t\right) /m\right] ^{1/2}$ is the
``thermal'' velocity and $f^{(0)}$ is an isotropic function of ${\bf V}%
^{\ast }$. Then Eq.\ (\ref{2.13aa}) is equivalent to 
\begin{equation}
J[f^{(0)},f^{(0)}]-\frac{1}{2}\zeta ^{(0)}\frac{\partial }{\partial {\bf V}}%
\cdot \left( {\bf V}f^{(0)}\right) =0.  \label{2.14}
\end{equation}

Consider the initial condition $f^{(1)}(0)=0$. This is a physically
interesting case for hydrodynamics since $f^{(1)}(0)$ does not contribute to
the initial value of the hydrodynamic fields, due to (\ref{2.12}), and $%
f^{(0)}(0)$ is entirely determined by the hydrodynamic initial values. In
this case, $f^{(1)}$ is manifestly proportional to the gradients of the
fields. Retaining only terms of first order in these gradients, the
Boltzmann equation becomes 
\begin{equation}
\left( \partial _{t}+\overline{L}\right) f^{(1)}=-f^{(0)}{\bf \Phi }_{\beta }%
{\bf \cdot \nabla }y_{\beta },  \label{2.17}
\end{equation}%
with the definitions 
\begin{equation}
{\bf \Phi }_{1}{\bf =}\frac{1}{n}\left( {\bf V}+\frac{T}{m}\frac{\partial }{%
\partial {\bf V}}\ln f^{(0)}\right) ,  \label{2.19}
\end{equation}%
\begin{equation}
\Phi _{2,ij}=\left( \frac{1}{d}\delta _{ij}{\bf V}\cdot \frac{\partial }{%
\partial {\bf V}}-V_{i}\frac{\partial }{\partial V_{j}}\right) \ln f^{(0)},
\label{2.20}
\end{equation}%
\begin{equation}
{\bf \Phi }_{3}=m^{-1}\frac{\partial }{\partial {\bf V}}\ln f^{(0)}-\frac{%
{\bf V}}{2T}\left( d+{\bf V}\cdot \frac{\partial }{\partial {\bf V}}\ln
f^{(0)}\right) .  \label{2.18}
\end{equation}%
A term from $\zeta \lbrack f]$ to first order in the gradients does not
occur since it vanishes. This follows from the fact that $\zeta $ is a
scalar and can only be proportional to $\nabla \cdot {\bf u}$ at this order.
Since $\Phi _{2,ij}({\bf V})$ is traceless, there is no such contribution.
The solution to Eq.\ (\ref{2.17}) can be written in the form 
\begin{equation}
f^{(1)}=-{\bf F}_{\beta }{\bf \cdot \nabla }y_{\beta }.  \label{2.21}
\end{equation}%
When Eq.\ (\ref{2.21}) is substituted into Eq.\ (\ref{2.17}), the terms from 
$\partial _{t}{\bf \nabla }y_{\beta }$ are higher order in the gradients,
except for $\beta =3$ which gives a contribution proportional to the cooling
rate. The coefficients of the gradients are then found to obey the equations 
\begin{equation}
\left( \partial _{t}+\overline{L}\right) {\bf F}_{1}-\frac{T\zeta ^{(0)}}{n}%
{\bf F}_{3}=f^{(0)}{\bf \Phi }_{1},  \label{2.29}
\end{equation}%
\begin{equation}
\left( \partial _{t}+\overline{L}\right) F_{2,ij}=f^{(0)}\Phi _{2,ij},
\label{2.31}
\end{equation}%
\begin{equation}
\left( \partial _{t}+\overline{L}-\frac{3}{2}\zeta ^{(0)}\right) {\bf F}%
_{3}=f^{(0)}{\bf \Phi }_{3}.  \label{2.30}
\end{equation}

\section{Scaling}

The apparent simplicity of Eqs.\ (\ref{2.29})-(\ref{2.31}) is misleading
since $\overline{L}$, $\zeta ^{(0)}$, and ${\bf \Phi }_{\alpha }$ are
functions of time through their dependence on the hydrodynamic fields. This
dependence occurs as well for a gas with elastic collisions, but it can
neglected in that case when solving the equations since it is proportional
to higher order gradients. For elastic collisions, the temperature gives a
time dependence that is of zeroth order in the gradients and cannot be
neglected. However, it can be removed by a change of variables to
dimensionless forms. A characteristic length scale is given \ by the mean
free path $\ell $ and a dimensionless time $s$ is defined accordingly by 
\begin{equation}
ds=\frac{v_{0}(t)}{\ell }dt.  \label{4.2}
\end{equation}%
For simplicity we will take here $\ell =(n\sigma ^{d-1})^{-1}$, where $%
\sigma $ is the diameter of the particles, omitting a factor that depends on
the dimension of the system. Integrating over an interval $t$ shows that $s$
is an average number of collisions during that time. Moreover, introduce
dimensionless functions ${\bf \Phi }_{\beta }^{\ast }$ and ${\bf F}_{\beta
}^{\ast }$ as 
\begin{equation}
{\bf \Phi }_{1}=\frac{2v_{0}(t)}{n}{\bf \Phi }_{1}^{\ast }\left( {\bf V}%
^{\ast }\right) ,\quad \Phi _{2,ij}=2\Phi _{2,ij}^{\ast }\left( {\bf V}%
^{\ast }\right) ,\quad \Phi _{3}=\frac{2}{mv_{0}(t)}{\bf \Phi }_{3}^{\ast }(%
{\bf V}^{\ast }),  \label{4.3}
\end{equation}%
\begin{equation}
{\bf F}_{1}=\frac{2\ell }{v_{0}^{d}(t)}{\bf F}_{1}^{\ast }\left( {\bf V}%
^{\ast },s\right) ,\quad F_{2,ij}=\frac{2n\ell }{v_{0}^{d+1}(t)}%
F_{2,ij}^{\ast }\left( {\bf V}^{\ast },s\right) ,\quad {\bf F}_{3}=\frac{%
2n\ell }{mv_{0}^{d+2}(t)}{\bf F}_{3}^{\ast }\left( {\bf V}^{\ast },s\right) .
\label{4.4}
\end{equation}%
In dimensionless form Eqs.\ (\ref{2.29})-(\ref{2.31}) become 
\begin{equation}
\left( \partial _{s}+\overline{{\cal L}}^{\ast }\right) \left( {\bf F}%
_{1}^{\ast }-{\bf F}_{3}^{\ast }\right) =f^{(0)\ast }\left( {\bf \Phi }%
_{1}^{\ast }-{\bf \Phi }_{3}^{\ast }\right) ,  \label{4.5}
\end{equation}%
\begin{equation}
\left( \partial _{s}+\overline{{\cal L}}^{\ast }+\frac{1}{2}\zeta ^{\ast
}\right) F_{2,ij}^{\ast }=f^{(0)\ast }\Phi _{2,ij}^{\ast },  \label{4.7}
\end{equation}%
\begin{equation}
\left( \partial _{s}+\overline{{\cal L}}^{\ast }-\frac{1}{2}\zeta ^{\ast
}\right) {\bf F}_{3}^{\ast }=f^{(0)\ast }{\bf \Phi }_{3}^{\ast }.
\label{4.6}
\end{equation}%
with the definitions 
\begin{equation}
\overline{{\cal L}}^{\ast }=\overline{L}^{\ast }+\frac{1}{2}\zeta ^{\ast }%
\frac{\partial }{\partial {\bf V}^{\ast }}\cdot {\bf V}^{\ast },  \label{4.8}
\end{equation}%
\begin{equation}
\overline{L}^{\ast }=\frac{\ell }{v_{0}(t)}\overline{L},\quad \zeta ^{\ast }=%
\frac{\ell \zeta ^{(0)}}{v_{0}(t)}\,.  \label{4.9}
\end{equation}%
The right sides of Eqs.\ (\ref{4.5})-(\ref{4.7}) are independent of $s$, as
are $\overline{{\cal L}}^{\ast }$ and $\zeta ^{\ast }$. The integration is
now trivial 
\begin{equation}
{\bf F}_{1}^{\ast }\left( {\bf V}^{\ast },s\right) ={\bf F}_{3}^{\ast
}\left( {\bf V}^{\ast },s\right) +\int_{0}^{s}ds^{\prime }e^{-s^{\prime }%
\overline{{\cal L}}^{\ast }}f^{(0)\ast }\left( {\bf \Phi }_{1}^{\ast }-{\bf %
\Phi }_{3}^{\ast }\right) ,  \label{4.10}
\end{equation}%
\begin{equation}
F_{2,ij}^{\ast }\left( {\bf V}^{\ast },s\right) =\int_{0}^{s}ds^{\prime
}e^{-s^{\prime }\left( \overline{{\cal L}}^{\ast }+\frac{1}{2}\zeta ^{\ast
}\right) }f^{(0)\ast }\Phi _{2,ij}^{\ast },  \label{4.12}
\end{equation}%
\begin{equation}
{\bf F}_{3}^{\ast }\left( {\bf V}^{\ast },s\right) =\int_{0}^{s}ds^{\prime
}e^{-s^{\prime }\left( \overline{{\cal L}}^{\ast }-\frac{1}{2}\zeta ^{\ast
}\right) }f^{(0)\ast }{\bf \Phi }_{3}^{\ast }.  \label{4.11}
\end{equation}%
In the derivation of the above expressions, the assumed initial condition,
implying that all the ${\bf F}_{\beta }^{\ast }$ vanish for $s=0$, has been
used.

\section{Constitutive equations}

The pressure tensor and heat flux are determined from their definitions in
Eqs.\ (\ref{2.7}) and (\ref{2.8}). Substitution of Eqs.\ (\ref{2.11}) and (%
\ref{2.21}) leads to results valid to first order in the gradients, 
\begin{equation}
P_{ij}[{\bf r},t|f]=p({\bf r},t)\delta _{ij}-\int d{\bf v}D_{ij}({\bf V})%
{\bf F}_{\beta }{\bf \cdot \nabla }y_{\beta },\ \quad  \label{5.1}
\end{equation}
\begin{equation}
{\bf q}[{\bf r},t|f]=-\int d{\bf v}\,{\bf S}({\bf V}){\bf F}_{\beta }{\bf %
\cdot \nabla }y_{\beta }.  \label{5.2}
\end{equation}
Using fluid symmetry, these expressions reduce to 
\begin{equation}
P_{ij}=p\delta _{ij}-\eta \left( \nabla_{j} u_{i}+\nabla_{i}u_{j} -\frac{2}{d%
}\delta _{ij} \nabla \cdot {\bf u} \right) ,  \label{5.3}
\end{equation}
\begin{equation}
\quad {\bf q}=-\kappa {\bf \nabla }T-\mu {\bf \nabla }n.  \label{5.4}
\end{equation}
These are the expected Navier-Stokes constitutive equations, except for the
additional term in the heat flux proportional to ${\bf \nabla }n$. It will
be seen below that this new term is due entirely to the deviation of $%
f^{(0)} $ from the Maxwellian. The expressions for the transport
coefficients are identified as 
\begin{equation}
\eta =\frac{2nm\ell v_{0}(t)}{d^{2}+d-2}\int d{\bf V}^{\ast }D_{ij}^{\ast }(%
{\bf V}^{\ast })F_{2,ij}^{\ast }\left( {\bf V}^{\ast },s\right),  \label{5.7}
\end{equation}
\begin{equation}
\kappa =\frac{n\ell v_{0}(t)}{d}\int d{\bf V}^{\ast }{\bf S}^{\ast }({\bf V}%
^{\ast })\cdot {\bf F}_{3}^{\ast }\left( {\bf V}^{\ast },s\right),
\label{5.6}
\end{equation}
\begin{equation}
\mu {\bf =}\frac{m\ell v_{0}^{3}(t)}{d}\int d{\bf V}^{\ast }{\bf S}^{\ast }(%
{\bf V}^{\ast })\cdot {\bf F}_{1}^{\ast }\left( {\bf V}^{\ast },s\right).
\label{5.5}
\end{equation}
The dimensionless \ forms of ${\bf S}$ and $D_{ij}$ are 
\[
{\bf S}^{\ast }\equiv \left( V^{\ast 2}-\frac{d+2}{2}\right) {\bf V}^{\ast },%
\hspace{0.3in}D_{ij}^{\ast }=V_{i}^{\ast }V_{j}^{\ast }-\frac{1}{d}V^{\ast
2}\delta _{ij}. 
\]

To put these expressions in the desired Green-Kubo form, define the adjoint
operators $L^{\ast}$ and ${\cal L}^{\ast}$ by 
\begin{equation}
\int d{\bf V}^{\ast }X({\bf V}^{\ast })\left( 
\begin{array}{c}
\overline{L}^{\ast } \\ 
\overline{{\cal L}}^{\ast }%
\end{array}
\right) Y({\bf V}^{\ast })=-\int d{\bf V}^{\ast } Y({\bf V}^{\ast}) \left( 
\begin{array}{c}
L^{\ast } \\ 
{\cal L}^{\ast }%
\end{array}
\right) X({\bf V}^{\ast }),  \label{5.8}
\end{equation}
for arbitrary functions $X$ and $Y$. From Eq.\ (\ref{4.8}) it is trivially
seen that 
\begin{equation}
{\cal L}^{\ast }=L^{\ast}+ \frac{1}{2}\zeta ^{\ast }{\bf V}^{\ast } \cdot 
\frac{\partial}{\partial {\bf V}^{\ast}}\, .  \label{5.9}
\end{equation}
The expression of $L^{\ast}$ can be obtained by using the properties of the
Boltzmann collision operator, but it will be not needed here.

Also, define ``correlation functions'' by 
\begin{equation}
\left\langle XY\right\rangle =\int d{\bf V}^{\ast }f^{(0)\ast }({\bf V}%
^{\ast })X({\bf V}^{\ast })Y\left( {\bf V}^{\ast }\right) .  \label{5.10}
\end{equation}
Then, with these definitions and Eqs.\ (\ref{4.10})-(\ref{4.11}), the
transport coefficients given by Eqs. (\ref{5.7})-(\ref{5.5}) can be
rewritten as 
\begin{equation}
\eta =\frac{2nm\ell v_{0}(t)}{d^{2}+d-2}\int_{0}^{s}ds^{\prime }\left\langle
D_{ij}^{\ast }(s^{\prime })\Phi _{2,ij}^{\ast }\right\rangle e^{-\frac{1}{2}%
s^{\prime} \zeta ^{\ast }},  \label{5.13}
\end{equation}
\begin{equation}
\kappa =\frac{n\ell v_{0}(t)}{d}\int_{0}^{s}ds^{\prime }\left\langle {\bf S}%
^{\ast }(s^{\prime })\cdot {\bf \Phi }_{3}^{\ast }\right\rangle e^{\frac{1}{2%
}s^{\prime}\zeta ^{\ast }},  \label{5.12}
\end{equation}
\begin{equation}
\mu {\bf =}\frac{2T\kappa}{n} +\frac{m\ell v_{0}^{3}(t)}{d}%
\int_{0}^{s}ds^{\prime }\left\langle {\bf S}^{\ast }(s^{\prime })\cdot
\left( {\bf \Phi }_{1}^{\ast }-{\bf \Phi }_{3}^{\ast }\right) \right\rangle .
\label{5.11}
\end{equation}
The time dependence of the correlation functions is given by 
\begin{equation}
X(s)=e^{s{\cal L}^{\ast }}X({\bf V}^{\ast }).  \label{5.14}
\end{equation}
Equations (\ref{5.11})-(\ref{5.13}) are the Green-Kubo expressions for the
transport coefficients of a low density granular gas. The spectrum of ${\cal %
L}^{\ast }$ is such that the correlation functions decay to zero for $s \gg
1 $. This sets the time scale for hydrodynamics. In this limit, the above
expressions agree with those obtained by the Chapman-Enskog method \cite%
{brey98boltz}.

\section{Discussion}

To discuss the GK expressions for a granular gas it is instructive to write
the corresponding results for elastic collisions $\left( \alpha =1\right) $.
In that case, $f^{(0)}$ is the local Maxwellian and 
\begin{equation}
{\bf \Phi }_{1}^{\ast }\rightarrow 0,\quad \Phi _{2,ij}^{\ast}\rightarrow
D_{ij}^{\ast }, \quad {\bf \Phi }_{3}^{\ast}\rightarrow {\bf S}^{\ast }({\bf %
V)}.  \label{6.1}
\end{equation}
Then $\mu \rightarrow \mu _{0} =0$, and 
\begin{equation}
\eta \rightarrow \eta _{0}=\frac{2n\ell v_{0}}{d^{2}+d-2} \int_{0}^{s}ds^{%
\prime }\left\langle D_{ij}^{\ast }(s^{\prime })D_{ij}^{\ast }\right\rangle,
\quad \kappa \rightarrow \kappa _{0}=\frac{n\ell v_{0}}{d}%
\int_{0}^{s}ds^{\prime } \left\langle {\bf S}^{\ast }(s^{\prime })\cdot {\bf %
S}^{\ast }\right\rangle .  \label{6.2}
\end{equation}
Since the temperature is constant to leading order in the gradients for $%
\alpha =1,$ the variable $s=\left(v_{0}/\ell \right) t$ is simply
proportional to time. These are the standard forms for the GK expressions in
terms of the autocorrelation functions composed from the ``microscopic''
fluxes ${\bf S}^{\ast }$ and $D_{ij}^{\ast }$. There are several differences
that occur for granular gases:

\begin{itemize}
\item The correlation functions are defined as averages over $f^{(0)}$. This
local HCS distribution is determined from (\ref{2.14}) and differs from the
Maxwellian for all $\alpha <1$.

\item The correlation functions are not constructed from ${\bf S}^{\ast }$
and $D_{ij}^{\ast }$ alone. Each is paired with another function from the
set ${\bf \Phi }_{\beta }^{\ast }$.

\item The time integration is replaced by an integration over the average
collision number, $s$. The correlation functions have approximate
exponential decay in the variable $s$ rather than $t$.

\item The integrals over $s$ are not controlled solely by the correlation
functions. In addition there are time dependent factors arising from the
change of the temperature over the duration of the integral.
\end{itemize}

The most surprising among these differences is the replacement of one of the
``microscopic'' fluxes ${\bf S}^{\ast }$ or $D_{ij}^{\ast }$ by one of the
new variables ${\bf \Phi }_{\beta }^{\ast }$ determined from $f^{(0)}$
through Eqs. (\ref{2.19})-(\ref{2.18}). If $f^{(0)}$ were replaced by the
local Maxwellian $f_{M}^{(0)\ast }$ in the above analysis, the correlation
functions $\left\langle {\bf S}^{\ast }(s^{\prime })\cdot {\bf S}^{\ast
}\right\rangle $ and $\left\langle D_{ij}^{\ast }(s^{\prime })D_{ij}^{\ast
}\right\rangle $ would appear in Eqs.\ (\ref{5.13})-(\ref{5.11}). However,
this would not be a consistent solution to the Boltzmann equation and the
transport coefficients would not have the correct dependence on $\alpha $
(e.g. $\mu $ would vanish for all $\alpha $). The role of $f^{(0)}$ is more
than just a local reference state with the exact moments for $n$, $T$, and $%
{\bf u}$. In addition, it must be an approximate solution for the dynamics.
In dimensionless form, $f^{(0)\ast }$ is a universal function of $V^{\ast }$
determined from the equation 
\begin{equation}
J^{\ast }[f^{(0)\ast },f^{(0)\ast }]-\frac{1}{2}\zeta ^{\ast }\frac{\partial 
}{\partial {\bf V}^{\ast }}\cdot \left( {\bf V}^{\ast }f^{(0)\ast }\right)
=0,\quad \zeta ^{\ast }=\zeta ^{\ast }[f^{(0)\ast }],  \label{6.3}
\end{equation}%
without reference to any particular hydrodynamic state. The above analysis
shows that (\ref{6.3}) is necessary for a consistent ordering of the
solution in terms of gradients. The local reference distribution for a given
hydrodynamic state follows from this solution according to 
\begin{equation}
f^{(0)}=n\left( {\bf r},t\right) \left( \frac{m}{2T\left( {\bf r},t\right) }%
\right) ^{d/2}f^{(0)\ast }\left[ \left( \frac{m}{2T\left( {\bf r},t\right) }%
\right) ^{1/2}\left( {\bf v}-{\bf u}\left( {\bf r},t\right) \right) \right] .
\label{6.4}
\end{equation}%
Its dependence on the hydrodynamic variables is determined by the solution
to (\ref{6.3}).

An estimate for the ${\bf \Phi }_{\beta }^{\ast }$ can be obtained from an
approximation to $f^{(0)\ast }$ obtained by a polynomial \ expansion \cite%
{brey96hcs,vanNoije98hcs}, which to leading order is%
\begin{equation}
f^{(0)\ast }=f_{M}^{(0)\ast }\left[ 1+\frac{1}{4}c^{\ast }(\alpha )\left(
V^{\ast 4}-(d+2)V^{\ast 2}+\frac{d(d+2)}{4}\right) \right],  \label{6.5}
\end{equation}%
\begin{equation}
c^{\ast }(\alpha )=\frac{32\left( 1-\alpha \right) \left( 1-2\alpha
^{2}\right) }{9+24d+(8d-41)\alpha+30\alpha^{2} (1-\alpha) }  \label{6.6}
\end{equation}%
This is known to be accurate to within a few percent for all $\alpha $ and $%
V^{\ast} \leq 1$. In this approximation the variables ${\bf \Phi }_{\beta
}^{\ast }$ become 
\begin{equation}
{\bf \Phi }_{1}^{\ast }{\bf =}\frac{1}{4}c^{\ast }(\alpha ){\bf S}^{\ast },
\label{6.7}
\end{equation}%
\begin{equation}
\Phi _{2,ij}^{\ast }=D_{ij}^{\ast }\left[ 1-\frac{1}{2}c^{\ast }(\alpha
)\left( V^{\ast 2}-\frac{d+2}{2}\right) \right] ,  \label{6.8}
\end{equation}%
\begin{equation}
{\bf \Phi }_{3}^{\ast }({\bf V}^{\ast })={\bf S}^{\ast }\left[ 1-\frac{1}{2}%
c^{\ast }(\alpha )\left( V^{\ast 2}-1\right) \right] .  \label{6.9}
\end{equation}

Linear response methods typically construct the response function for
spatial perturbation of the \ homogeneous state of a specific type: those
that couple only to the microscopic conserved densities. This follows from
consideration of an initial local equilibrium state or maximum entropy
state. Then manipulation of the response function for the conserved
densities using the microscopic conservation laws leads to GK expressions in
terms of the microscopic fluxes. However, it is clear that expansion of (\ref%
{6.4}) for small spatial perturbations does not give a representation in
terms of the conserved densities. Thus, a straightforward application of
linear response can lead to formal results with internal inconsistencies.
One such inconsistency can be a failure of the solubility conditions for the
linear integral equations associated with (\ref{4.5}) - (\ref{4.7}). In the
GK form these conditions translate to requirements that the fluxes be
orthogonal to invariants of the dynamics. These issues and a derivation of
the GK expressions from the Liouville equation will be given elsewhere.

\section{Acknowledgments}

The research of JWD was supported in part by the Department of Energy grant
(DE-FG03-98DP00218). The research of JJB was partially supported by the
Direcci\'{o}n General de Investigaci\'{o}n Cient\'{\i}fica y T\'{e}cnica
(Spain) through Grant No. PB98-1124.


\bigskip

\begin{references}
\bibitem{mclennan89} J. A. McLennan, {\em Introduction to Nonequilibrium
Statistical Mechanics} (Prentice-Hall, New Jersey, 1989).

\bibitem{brey97statmech} J. J. Brey, J. W. Dufty, and A. Santos,
``Dissipative Dynamics for Hard Spheres'', J. Stat. Phys. {\bf 87}, 1051
(1997).

\bibitem{dufty00granada} J. W. Dufty, ``Statistical mechanics, kinetic
theory, and hydrodynamics for rapid granular flow'', J. Phys.: Condens.
Matter 12, A {\bf 47 }(2000).

\bibitem{brey01rev} J. J. Brey and D. Cubero, ``Hydrodynamic transport
coefficients of granular gases'' in {\em Granular Gases}, eds. T. Poschel
and S. Luding (Springer, NY, 2001).

\bibitem{goldhirsh01} I. Goldhirsh, ``Granular gases - probing the
boundaries of hydrodynamics'', in {\em Granular Gases}, eds. T. Poschel and
S. Luding (Springer, NY, 2001).

\bibitem{vanNoije01sm} T. P. C. van Noije and M. H. Ernst, ``Kinetic theory
of granular gases'', in {\em Granular Gases}, eds. T. Poschel and S. Luding
(Springer, NY, 2001).

\bibitem{brey98boltz} J. J. Brey, J. W. Dufty, C. S. Kim, A. Santos,
``Hydrodynamics for Granular Flow at Low Density'', Phys. Rev. E {\bf 58},
4638 (1998).

\bibitem{garzo00enskog} V. Garz\'{o} and J. Dufty, ``Dense fluid transport
for inelastic hard spheres'', Phys. Rev. E {\bf 59}, 5895 (1999).

\bibitem{garzo01mix} V. Garz\'{o} and J. Dufty, \ ``Hydrodynamics for a
granular mixture at low density'', Phys. Fluids (in press); cond-mat/0105395
v1.

\bibitem{goldhirsh00GK} I. Goldhirsh and T. P. C. van Noije, ``Green-Kubo
relations for granular fluids'' Phys. Rev. E {\bf 61}, 3241 (2000).

\bibitem{brey96hcs} J. J. Brey, M. J. Ruiz-Montero, and D. Cubero,
``Homogeneous cooling state of a low-density granular flow'', Phys. Rev. E 
{\bf 54}, 3664 (1997).

\bibitem{vanNoije98hcs} T. P. C. van Noije and M. H. Ernst, ``Velocity
distributions in homogeneous granular fluids: the free and heated case'',
Granular Matter {\bf 1}, 57 (1998).

\bibitem{dufty02} J. W. Dufty, J. F. Lutsko, and J. J. Brey, unpublished.

\bibitem{dufty01mob} J. W. Dufty and V. Garzo, ``Mobility and diffusion in
granular flow'', J. Stat. Phys. {\bf 105}, 723 (2001).

\bibitem{dufty02trieste} J. W. Dufty, ``Kinetic Theory and Hydrodynamics for
a Low Density Gas'', J. W. Dufty, Advances in Complex Systems (in press);
cond-mat/0109215.

\bibitem{dorfman} J. R. Dorfman and H. van Beijeren, ``The kinetic theory of
gases'' in {\em Statistical Mechanics, Part B}, edited by B. Berne (Plenum
Press, NY, 1977).
\end{references}
\end{document}